\documentclass[prc,aps,twocolumn]{revtex4}
\usepackage{graphicx}
\usepackage{dcolumn}
\usepackage{bm}

\def\tend{\mathop{\to}}

\def\lim{\mathop{\rm {lim}}}

\begin{document}
\draft \preprint{HEP/123-qed}
\title{Nonlocality of nucleon interaction and \\
an anomalous off shell behavior of the two-nucleon amplitudes}
\author{Renat Kh.Gainutdinov and Aigul A.Mutygullina}
\address{
Department of Physics, Kazan State University, 18 Kremlevskaya St,
Kazan 420008, Russia } \email{Renat.Gainutdinov@ksu.ru}

\begin{abstract}
The problem of the ultraviolet divergences that arise in
describing the nucleon  dynamics at low energies is considered. By
using the example of an exactly solvable model it is shown that
after renormalization the interaction generating nucleon dynamics
is nonlocal in time. Effects of such nonlocality on low-energy
nucleon dynamics are investigated. It is shown that nonlocality in
time of nucleon-nucleon interactions gives rise to an anomalous
off-shell behavior of the two-nucleon amplitudes that have
significant effects on the dynamics of many-nucleon systems.
\end{abstract}
\maketitle \narrowtext

\section{Introduction}
\label{sec:level1}

Investigations aimed at assessing the extent to which quarks and
gluons bound in hadrons can affect low-energy nucleon dynamics are
of great importance for obtaining deeper insights into the nature
of strong interactions. These fundamental degrees of freedom
manifest themselves, for example, as symmetries in low-energy
nucleon-nucleon interaction ($NN$) that are compatible with QCD
symmetries. In the most natural way, the symmetries in question
are taken into account within an effective field theory [1], which
is extensively used at present in describing low-energy nucleon
dynamics. Quark and gluon degrees of freedom also manifest
themselves in that the interaction of nucleons must be nonlocal in
time because of the presence of these intrinsic degrees of
freedom. Accordingly, the effective potentials of $NN$ interaction
must be energy-dependent. The possibility of using such potentials
in describing hadron-hadron interactions at low and intermediate
energies was extensively discussed in the literature [2]. It may
be seem that this time nonlocality of the effective operator of
$NN$ interaction is not compatible with an effective field theory,
which is a local theory. However, this is not so. Indeed, an
effective field theory leads to effective $NN$- interaction
operator whose ultraviolet behavior is "bad"; that is, matrix
elements as functions of momenta decrease at infinity
insufficiently fast for the Schr{\"o}dinger equations to be
meaningful. For this reason, it is necessary to regularize these
equations and to renormalize the potentials. Ultraviolet
divergences stem from locality of the theory; that is, they are
due to the disregard of the fact that $NN$ interaction cannot be
local because of the presence of intrinsic quark and gluon degrees
of freedom.

As a matter of fact, we run here into the same problem as in
 quantum field theory: locality of the theory leads to
ultraviolet divergences, but the introduction of a nonlocal form
factor in the Hamiltonian or in the interaction
 Lagrangian violates the covariance of the theory. The reason behind
this is quite obvious. The Schr{\"o}dinger equation is local in
time,  and the  Hamiltonian describes an instantaneous
 interaction; in relativistic
  theory a process that is local-in-time  must be local in space as well.
For the introduction of a nonlocality in a theory be
 self-consistent, it is necessary to extend quantum dynamics to the case
of the evolution of quantum systems whose  dynamics is governed by
an interaction that is nonlocal-in-time.
 For the first time, this problem was solved
in [3], where it was shown that the simultaneous use of basic
principles of the canonical and the Feynman formulation of quantum
theory opens the possibility for generalizing quantum dynamics in
this way. The generalized dynamical equation derived in [3] by
using only generally accepted principles of quantum theory makes
it possible to describe the evolution of quantum systems not only
for the case where the fundamental interaction is instantaneous
(it then reduces to the Schr{\"o}dinger equation) but also in the
case where the interaction is nonlocal in time. It was shown in
[3] that generalized quantum dynamics developed in this way opens
new possibilities for solving the problem of ultraviolet
divergences in quantum field theory. An exactly solvable model was
constructed in [4,5] for investigating the character of the
dynamics of quantum systems controlled by an interaction that is
nonlocal in time and was used, by way of example [3], to show that
there is one-to-one correspondence between the ultraviolet
behavior of the model form factors and the nonlocality of the
interaction. If the high -momentum behavior of the form factors
satisfies the usual requirements of the Hamiltonian formalism, the
interaction in the system is inevitably local, but, if this is not
so (that is, the behavior of the form factors leads to ultraviolet
divergences in Hamiltonian dynamics), the interaction in the
system can only be nonlocal. In the latter case, the form of the
nonlocal interaction operator is unambiguously determined by the
asymptotic high-momentum behavior of the form factors, the
dynamics of the system not being Hamiltonian here.

In connection with the fact that effective field theories lead to
models where the effective potentials of $NN$ interaction exhibit
a "bad" ultraviolet behavior (see above), interest in studying
such models - in particular, in their regularization and
renormalization - has been quickened in recent years. For example,
the problem of a dimensional regularization of the Lippmann-
Schwinger equation was studied in [6] by considering the example
of a model where $NN$ interaction is described by a separable
potential featuring a form factor that leads to a logarithmic
singularity. In that study, the application of the regularization
and renormalization procedure to the coupling constant made it
possible to obtain the $T$ matrix of the nonlocal model proposed
in [4,5] for the corresponding form factor. Thus, it was  found
that, upon the renormalization, the effective $NN$ interaction,
which governs the dynamics of the system being considered, becomes
nonlocal in time, the evolution of the system being described by a
dynamical equation that is not equivalent to the Schr{\"o}dinger
equation, but which is an equation of the type associated with
generalized quantum dynamics. It should be emphasized that, in the
case being discussed, generalized quantum dynamics permits
treating the evolution of the system as rigourously as this is
done in the case where the ultraviolet behavior of the form
factors is such that the dynamics of the system is Hamiltonian. A
constriction of the model in question by applying the
renormalization method only makes it possible to determine the
two-nucleon $T$ matrix, but it gives no way to derive an equation
that would describe nucleon dynamics. The latter in turn prevents
the use of these results in describing the dynamics of
multinucleon systems. At the same time, the theory of
renormalizations can provide the possibility of constructing, on
the basis of an effective field theory, an effective
$NN$-interaction operator that is nonlocal in time and which is
compatible with its symmetries. This operator can then be employed
to describe nucleon dynamics in terms of the dynamical equation of
generalized quantum dynamics. This may open new possibilities for
developing the theory of $NN$ interactions that is based on the
effective field theory. Needless to say, realistic models to which
the effective field theory must lead will be more involved than
the model considered in [4,5]. As we have already mentioned, this
exactly solvable model reflects, however, a crucial feature of the
$NN$ interaction - namely, the bad ultraviolet behavior of matrix
elements as functions of momenta, which takes place if the
interaction is nonlocal in time. In the present study, we address
the problem of assessing the extent to which the nonlocality of
the $NN$ interaction in time can affect the character of nucleon
dynamics. We will show that this nonlocality of the $NN$
interaction leads to an anomalous off-shell behavior of
two-nucleon amplitudes, which has a pronounced effect on the
dynamics of multinucleon systems.

\section{Generalized quantum dynamics}

 The basic
concept of the canonical formalism of quantum theory is that it
can be formulated in terms of vectors of a Hilbert space and
operators acting on this space. The postulates  establish the
connection between the vectors and operators and states of a
quantum system and observables.  In the canonical formalism  these
postulates are used together with the dynamical postulate
according to which the time evolution of a quantum system is
governed by the Schr{\"o}dinger equation. However, this equation
permits only instantaneous interactions. At the same time, in
Ref.[3] it has been shown that this equation is not most general
dynamical equation consistent with the current concepts of quantum
physics. It has been shown [3] that the use of the above basic
postulates of canonical formalism in combination with the basic
postulate of the Feynman approach according to which the
probability amplitude of an event that can be happen in several
different ways is a sum of the probability amplitudes for each to
these ways gives rise to a more general dynamical equation.

From the postulates of the canonical formalism it follows that the
time evolution of a quantum system is described by the evolution
operator that must be unitary
\begin{equation}
U^{+}(t,t_0) U(t,t_0) = U(t,t_0) U^{+}(t,t_0) = {\bf 1},
\end{equation}
and satisfies the composition low
\begin{equation}
 U(t,t') U(t',t_0) = U(t,t_0),
\quad U(t_0,t_0) ={\bf 1}.
\end{equation}
The evolution operator in the interaction picture can be written
in the form
\begin{eqnarray}
\langle\psi_2| U(t,t_0)|\psi_1\rangle = \langle\psi_2|\psi_1\rangle +\nonumber\\
+\int_{t_0}^t dt_2 \int_{t_0}^{t_2} dt_1 \langle\psi_2|\tilde
S(t_2,t_1)|\psi_1\rangle, \label{repre}
\end{eqnarray}
where $\langle\psi_2|\tilde S(t_2,t_1)|\psi_1\rangle$ is the
probability amplitude that if at time $t\to-\infty$ the system was
in the state $|\psi_1\rangle,$ then the interaction in the system
will begin at time $t_1$ and will end at  time $t_2,$ and at time
$t\to\infty$ the system will be in the state $|\psi_2\rangle.$ The
first term on the right-hand side of (3) corresponds to the
evolution process that is free from interaction. Indeed, according
to the basic postulates of Feynman formalism the probability
amplitude describing the matrix element
$\langle\psi_2|U(t,t_0)|\psi_1\rangle$ can be represented as a sum
of contributions from all alternative ways of realization of the
corresponding evolution process and the amplitudes
$\langle\psi_2|\tilde S(t_2,t_1)|\psi_1\rangle$ is contributions
to the amplitude $\langle\psi_2|U(t,t_0)|\psi_1\rangle$ of such
alternatives.

As has been shown in Ref.[3], for the evolution operator
$U(t,t_0)$ given by (\ref{repre}) to be unitary, the operator
$\tilde S(t_2,t_1)$ must satisfy the following equation:
\begin{eqnarray}
(t_2-t_1) \tilde S(t_2,t_1) = \int^{t_2}_{t_1} dt_4
\int^{t_4}_{t_1}dt_3 \nonumber \\
 \times(t_4-t_3) \tilde S(t_2,t_4) \tilde S(t_3,t_1).
\label{main}
\end{eqnarray}
This equation allows one to obtain the amplitudes
$\langle\psi_2|\tilde S(t_2,t_1)|\psi_1\rangle$ for any $t_1$ and
$t_2$, if the amplitudes $\langle\psi_2|\tilde S(t'_2,
t'_1)|\psi_1\rangle$ corresponding to infinitesimal duration times
$\tau = t'_2 -t'_1$ of interaction are known. It is natural to
assume that most of the contribution to the evolution operator in
the limit $t_2 \to t_1$ comes from the processes associated with
the fundamental interaction in the system under study. Denoting
this contribution by $H_{int}(t_2,t_1)$, we can write
\begin{equation}
\tilde{S}(t_2,t_1) = H_{int}(t_2,t_1) + \tilde S_1(t_2,t_1),
\label{eqbound}
\end{equation}
where $\tilde S_1(t_2,t_1)$ is the part of the operator $\tilde
S(t_2,t_1)$ which in the limit $t_2\to t_1$ gives a negligibly
small contribution to the evolution operator in comparison with
$H_{int}(t_2,t_1)$. We will assume that the operator
$H_{int}(t_2,t_1)$ contain all the dynamical information that is
needed to construct the evolution operator. From the mathematical
point of view the requirement that the operator $H_{int}(t_2,t_1)$
must have such a form that the Eq.(4) has a unique solution having
the following behavior near the point $t_2=t_1$:
\begin{equation}
\tilde{S}(t_2,t_1) \tend\limits_{t_2\rightarrow t_1}
H_{int}(t_2,t_1) + o(\tau^{\epsilon}), \label{bound}
\end{equation}
where $\tau=t_2-t_1$ and the value of the parameter $\epsilon$
depends on the form of the operator $ H_{int}(t_2,t_1).$

The operator $H_{int}(t_2,t_1)$ plays the role which the
interaction Hamiltonian plays in the ordinary formulation of
quantum theory: It generates the dynamics of a system. Being a
generalization of the interaction Hamiltonian, this operator is
called the generalized interaction operator. If $H_{int}(t_2,t_1)$
is specified, Eq.(\ref{main}) allows one to find the operator
$\tilde S(t_2,t_1).$ Representation for the evolution operator
(\ref{repre}) can then be used to construct the evolution operator
$U(t,t_0)$ at any time $t$ and $t_0$. Thus Eq.(\ref{main}) can be
regarded as an equation of motion for states of a quantum system
and it is used as a basic dynamical equation.

The equation of motion (\ref{main}) is equivalent to the following
differential equation:
\begin{equation}
\frac{d T(z)}{dz} =- \sum \limits_{n}\frac{T(z)|n\rangle\langle n
|T(z)}{(z-E_n)^2}, \label{dif}
\end{equation}
where the operator $T(z)$ is defined by
\begin{eqnarray}
\langle n_2|T(z)|n_1\rangle \\
= i \int_{0}^{\infty} d\tau \exp(iz\tau) \langle n_2|\tilde
T(\tau)|n_1\rangle. \nonumber\label{tt}
\end{eqnarray}
Here $|n\rangle$ are the eigenvectors of the free Hamiltonian:
$H_0|n\rangle=E_n|n\rangle$ and $n$ stands for the entire set of
discrete and continuous variables that characterize the system in
full. According (6) and (8), the boundary condition for Eq.(7) has
the form
\begin{equation}
\langle n_2|T(z)|n_1\rangle \tend \limits_{|z| \tend \infty}
\langle n_2| B(z)|n_1\rangle + o(|z|^{-\beta}),\label{dbound}
\end{equation}
where $\beta=1+\epsilon,$ and
\begin{equation}
B(z) = i \int_0^{\infty} d\tau \exp(iz \tau) H^{(s)}_{int}(\tau),
\end{equation}
with
 $$H^{(s)}_{int}(t_2-t_1) =
 \exp(-iH_0t_2)H_{int}(t_2,t_1)\exp(iH_0t_1).$$
The operator $B(z)$, that it was called an effective interaction
operator, must be so close to the solution of Eq.(7) in the limit
$|z|\to\infty$, that this differential equation has an unique
solution having the asymptotic behavior (9).

 The dynamics governed
by Eq.(\ref{main}) is equivalent to the Hamiltonian dynamics in
the case where the generalized interaction operator  is of the
form
\begin{equation}
 H_{int}(t_2,t_1) = - 2i \delta(t_2-t_1)
 H_{I}(t_1) ,
 \label{delta}
\end{equation}
$H_{I}(t_1)$ being the interaction Hamiltonian in the interaction
picture. In this case one can derive the Schr{\"o}dinger equation
from Eq.(4).  The delta function $\delta(\tau)$ in (\ref{delta})
emphasizes that in this case the fundamental interaction is
instantaneous. Thus the Schr{\"o}dinger equation results from the
generalized equation of motion (\ref{main}) in the case where the
interaction generating the dynamics of a quantum system is
instantaneous. At the same time, Eq.(\ref{main}) permits the
generalization to the case where the operator $H_{int}(t_2,t_1)$
has no such a singularity as the delta function at the point
$t_2=t_1$. In this case the fundamental interaction generating the
dynamic of a quantum system is nonlocal in time: The evolution
operator is determined the generalized interaction operator
$H_{int}(t_2,t_1)$ as a function of a duration time of interaction
$\tau=t_2-t_1$. Below we will demonstrate this fact by using the
example of exactly solvable models.

\section{Models with nonlocal in time interactions and nucleon dynamics}

Let us consider the evolution problem for two nucleons in the
c.m.s. We denote the relative momentum by ${\bf p}$ and the
reduced mass by $\mu.$ Assume that the generalized interaction
operator in the Schr{\"o}dinger picture $H^{(s)}_{int}(\tau)$ has
the form
\begin{equation}
\langle{\bf p}_2| H^{(s)}_{int}(\tau) |{\bf p}_1\rangle =
\psi^*({\bf p}_2) \psi({\bf p}_1) f(\tau),
\end{equation}
where $f(\tau)$ is some function of $\tau$.  Let the form factor
$\psi ({\bf p})$ be of the form
\begin{equation}
\psi({\bf p}) = |{\bf p}|^{-\alpha}+g({\bf {p}}),
\end{equation}
and in the limit $|{\bf p}| \tend \infty$ the function $g({\bf
{p}})$ satisfies the estimate $g({\bf {p}})=o(|{\bf
{p}}|^{-\delta})$, where $\delta>\alpha,$ $\delta>\frac{3}{2}.$
The solution  $\langle{\bf p}_2| T(z) |{\bf p}_1\rangle$ is of the
form
$$
  \langle{\bf p}_2| T(z)|{\bf p}_1\rangle = \psi^* ({\bf p}_2)\psi ({\bf p}_1)
t(z),\label{separ}
$$
 where, as it follows from (\ref{dif}), the function $t(z)$
satisfies the equation
\begin{equation}
\frac {dt(z)}{dz} = -t^2(z) \int d^3k \frac {|\psi ({\bf k})|^2}
{(z-E_k)^2} \label{deq}
\end{equation}
with the asymptotic condition
\begin{equation}
t(z)  \tend \limits_{|z| \tend \infty} f_1(z) + o(|z|^{-\beta}),
\label{asym}
\end{equation}
where $ f_1(z)= i \int_0^{\infty} f(\tau) \exp(iz\tau)d\tau , $
 and   the value of $\beta$ depends on  the form of the generalized
 interaction operator and it is determined by the condition that the solution
 of the differential equation must be unique.

 The solution of Eq.(\ref{deq}) with the
initial condition $t(a)=g_a,$ where $a \in (-\infty,0),$ is
\begin{eqnarray}
t(z) = g_a \\
\times\left(1 +(z-a) g_a
 \int d^3k \frac {|\psi ({\bf k})|^2}
{(z-E_k)(a-E_k)} \right)^{-1}.\nonumber\label{deqa}
\end{eqnarray}

If $\alpha >\frac{1}{2}$, in which case the form factors
$\psi({\bf p})$ satisfy the usual requirements of quantum
mechanics, the function $t(z)$ tends to a constant for $|z| \tend
\infty$,
\begin{equation}
t(z)  \tend \limits_{|z| \tend \infty} \lambda;\label{lambda}
\end{equation}
that is $f_1(z)=\lambda$. From this it follows that the only
possible form of the function $f(\tau)$ is $ f(\tau) = -2i \lambda
\delta(\tau) + f^{\prime}(\tau),$ where the function
$f^{\prime}(\tau)$ has no such a singularity at the point $\tau=0$
as the delta function. In this case  the generalized interaction
operator $H^{(s)}_{int}(\tau)$ has the form
\begin{equation}
\langle{\bf p}_2| H^{(s)}_{int}(\tau) |{\bf p}_1\rangle=-2i
\lambda \delta(\tau)\psi^*({\bf p}_2) \psi({\bf
p}_1),\label{instant}
\end{equation}
 and hence the dynamics generated by this operator is
equivalent to the dynamics governed by the Schr{\"o}dinger
equation with the separable potential $ \langle{\bf p}_2|H_I|{\bf
p}_1\rangle = \lambda \psi^*({\bf p}_2) \psi({\bf p}_1). $ Indeed,
solving Eq.(\ref{deq}) with the boundary condition (\ref{lambda}),
we easily get the well-known expression for the $T$ matrix in the
separable-potential model
\begin{eqnarray}
\langle{\bf p}_2|T(z)|{\bf p}_1\rangle\\
 = \lambda \psi^* ({\bf
p}_2)\psi({\bf p}_1) \left (1 - \lambda \int d^3k \frac
{|\psi({\bf k})|^2}{z-E_k} \right )^{-1} \nonumber\label{sol}
\end{eqnarray}

Standard quantum mechanics does not permit the extension of the
above model to the case $\alpha \leq \frac{1}{2}$ in (13). Indeed,
in the case of such a large-momentum behavior of the form factors
$\psi({\bf p}),$ the use of the interaction Hamiltonian given by
(18) leads to the UV divergences, i.e. the integral in (\ref{sol})
is not convergent. We will now show that the generalized dynamical
equation (\ref{main}) allows one to extend this model to the case
$-\frac{1}{2} <\alpha <\frac{1}{2}.$ Let us determine the class of
the functions $f_1(z)$ and correspondingly the value of $\beta$
for which Eq.(\ref{deq}) has a unique solution having the
asymptotic behavior (\ref{asym}). In the case $\alpha
<\frac{1}{2},$ the function $t(z)$ given by (\ref{deqa}) has the
following behavior for $|z| \tend \infty:$
\begin{widetext}
\begin{eqnarray}
\left\{
\begin{array}{rcl}
t(z)  \tend \limits_{|z| \tend \infty}  b_1(\alpha)
(-z)^{\alpha-\frac{1}{2}}+ b_2(\alpha) (-z)^{2 \alpha-1}+
o(|z|^{2 \alpha-1}),\quad -\frac{1}{2}<\alpha<\frac{1}{2};\\
t(z)  \tend \limits_{|z| \tend \infty}b_1(\alpha)\ln^{-1}(-z)+
b_2(\alpha)\ln^{-2}(-z)
+o(\ln^{-2}(-z)), \quad \alpha=\frac{1}{2}.\\
\end{array}
\right.
\end{eqnarray}
where
\begin{equation}
\left\{
\begin{array}{rcl}
b_1(\alpha) =- \frac{1}{2} \cos(\alpha \pi) \pi^{-2}
(2 \mu)^{\alpha-\frac{3}{2}},\\
b_2(\alpha)= b_1(\alpha) |a|^{\frac{1}{2}-
\alpha} -b_1^2(\alpha)(M(a)+g_a^{-1}),\quad -\frac{1}{2}<\alpha<\frac{1}{2};\\
b_1(\frac{1}{2})=-(4\pi \mu)^{-1},\quad
b_2(\frac{1}{2})= b_1(\frac{1}{2}) \ln(-a) -b_1^2(\frac{1}{2})(M(a)+g_a^{-1}),\\
\end{array}
\right.
\end{equation}
\end{widetext}
\begin{equation}
M(a) = \int \frac {|\psi({\bf k})|^2-|{\bf {k}}|^{-2\alpha}}
{a-E_k}d^3k.
\end{equation}
It should be emphasized that, in conventional quantum mechanics,
the vanishing of the $T$ matrix at infinity implies the vanishing
of the potential. From the point of view of standard theory, this
case is therefore trivial: there is no scattering in the system.
On the other hand, we have shown above that, in the case of
$\alpha\leq1/2$ in (13), Eq. (14) has a nontrivial solution that
tends to zero for $|z|\to\infty$, the dynamics here being governed
by the character of the vanishing of the $T$ matrix for
$|z|\to\infty$ rather than by its value at infinity. It can easily
be proven that all integral curves of the differential Eq. (14)
have the same first term, differing only by the values of the
parameter $b_2(\alpha)$. In order to obtain a unique solution of
Eq. (14), we therefore have to determine the first two terms in
the asymptotic expansion of $t(z)$ for $|z|\to\infty$, whence it
follows that the function $f_1(z)$ must have the form
\begin{widetext}
$$
\left\{
\begin{array}{rcl}
f_1(z) = b_1(\alpha) (-z)^{\alpha-\frac{1}{2}} +
b_2(\alpha) (-z)^{2 \alpha -1},\quad -\frac{1}{2}<\alpha<\frac{1}{2};\\
f_1(z) = b_1(\alpha) \ln^{-1}(-z) +
b_2(\alpha) \ln^{-2}(-z),\quad \alpha=\frac{1}{2},\\
\end{array}
\right.
$$
\end{widetext}
where $b_1(\alpha)$ is determined by (21), and only the parameter
$b_2(\alpha)$ is arbitrary. However, if there is a bound state in
the channel under study, then the parameter $b_2(\alpha)$ is
completely determined by demanding that the $T$ matrix has the
pole at the bound-state energy. For example, in the ${}^3S_1$
channel the $T$ matrix has a pole at energy $E_B=-2.2246$MeV. This
means that $\left[t(E_B)\right]^{-1}=0$, and, by putting $a=E_B$
in Eq.(\ref{deqa}), we get
\begin{equation}
\left [ t(z)\right ]^{-1}=(z-E_B)\int d^3 k\frac{|\psi({\bf
{k}})|^2}{(z-E_k) (E_B-E_k)}.\label{our}
\end{equation}
In this case, the parameter $b_2(\alpha)$ will accordingly have
the value
\begin{widetext}
$$
\left\{
\begin{array}{rcl}
b_2(\alpha)= b_1(\alpha)(-E_c)^{\frac{1}{2}-
\alpha} -b_1^2(\alpha)M(E_c),\quad -\frac{1}{2}<\alpha<\frac{1}{2};\\
b_2(\alpha)=b_1(\alpha)\ln(-E_c)-b_1^2(\alpha)M(E_c),\quad \alpha=\frac{1}{2}.\\
\end{array}
\right.
$$
Taking into account, that
$f(\tau)=\frac{i}{2\pi}\int_{-\infty}^{\infty}exp(-iz\tau)
f_1(z)dz,$ for the generalized interaction operator, we get
\begin{equation}
\left\{
\begin{array}{rcl}
\langle{\bf p}_2| H^{(s)}_{int}(\tau)|{\bf p}_1\rangle =
 \psi ({\bf p}_2)\psi^* ({\bf p}_1) \left(a_1\tau^{-\alpha-\frac{1}{2}} +
a_2\tau^{-2 \alpha}\right),\quad -\frac{1}{2}<\alpha<\frac{1}{2};\\
\langle{\bf p}_2| H^{(s)}_{int}(\tau)|{\bf p}_1\rangle = \psi
({\bf p}_2)\psi^* ({\bf p}_1)\frac{i}{2\pi}
\int_{-\infty}^{\infty}exp(-iz\tau)
\left(\frac{b_1(\alpha)}{\ln(-z)}+\frac{b_2(\alpha)}{\ln^2(-z)}
\right)dz, \quad \alpha=\frac{1}{2},\\
\end{array}
\right.
\end{equation}
where $a_1= -ib_1(\alpha) \Gamma ^{-1}(\frac{1}{2}-\alpha)
exp[i(-\frac{\alpha}{2}+ \frac{1}{4}) \pi],$ ¨ $a_2= -b_2(\alpha)
\Gamma ^{-1}(1-2\alpha) exp(-i \alpha \pi).$ By using (16), it is
easy to show that the corresponding solution for the $T$ matrix
has the form
\begin{equation}
\langle{\bf p}_2| T(z)|{\bf p}_1\rangle = N(z)\psi ({\bf
p}_2)\psi^* ({\bf p}_1),
\end{equation}
where $$ \left\{
\begin{array}{rcl}
N(z)=b_1^2(\alpha)[-b_2(\alpha)+b_1(\alpha)(-z)^{\frac{1}{2}-\alpha}
+M(z)b_1^2(\alpha)]^{-1}, \quad -\frac{1}{2}<\alpha<\frac{1}{2};\\
N(z)=b_1^2(\alpha)[-b_2(\alpha)+b_1(\alpha) \ln(-z)
+M(z)b_1^2(\alpha)]^{-1}, \quad \alpha=\frac{1}{2}.\\
\end{array}
\right.
$$
\end{widetext}
Taking into account that the $T$ matrix is connected with the
evolution operator by the equation [3]
\begin{eqnarray}
U(t,t_0) = {\bf 1} + \frac {i}{2\pi}  \\
\times\int_{-\infty}^{\infty} dx \frac{\exp[-i(z-H_0)t]}{(z-H_0)}
T(z)\frac{\exp[i (z - H_0)t_0]}{(z-H_0)},\nonumber
\end{eqnarray}
where  $z=x+iy$, $y>0$, for the evolution operator, we get
\begin{eqnarray}
\langle{\bf p}_2|U(t,t_0)|{\bf p}_1\rangle = \langle{\bf p}_2|{\bf
p}_1\rangle + \frac
{i}{2\pi} \int_{-\infty}^{\infty} dx\nonumber\\
\times \frac {\exp[-i(z-E_{p_2})t] \exp[i (z - E_{p_1})t_0]}
{(z-E_{p_2})(z-E_{p_1})} \nonumber\\
\times N(z)\psi ({\bf p}_2)\psi^* ({\bf p}_1).
\end{eqnarray}
By using (27), we can then construct a vector representing the
state of the system at any instant of time $t$. It can be shown
that the evolution operator (27) is unitary if the parameter
$b_2(\alpha)$ is real-valued and that it satisfies the composition
law.

It should also be noted that there is the one-to-one
correspondence between the behavior of the form factors $\psi({\bf
p})$ for $|{\bf p}|\to\infty$ and the character of dynamics: if
the form factors satisfy the usual requirements of quantum
mechanics [for $\alpha>\frac{1}{2}$ in the asymptotic expression
(13)], the generalized interaction operator must have the form
(18). In this case, the fundamental interaction is instantaneous.
At $\alpha\leq1/2$, in which case the high-momentum behavior of
the form factors in Hamiltonian dynamics leads to ultraviolet
divergences, the only possible form of $H^{(s)}_{int}(\tau)$ is
that which is given by (24); that is, the fundamental interaction
that generates dynamics in the quantum system being considered is
nonlocal in time.

We have shown that generalized quantum dynamics makes it possible
to describe, in a natural way, the evolution of quantum systems
where the interaction leads to ultraviolet divergences in
Hamiltonian dynamics. It was indicated above that, in effective
field theories, one has to deal with precisely such interactions,
with the result that it becomes necessary to invoke various
regularization and renormalization procedures. In [6], this
problem was investigated for the example where the $NN$
interaction is described by the
 separable potential
$\langle{\bf p}_2|V|{\bf p}_1\rangle=\lambda\psi^*({\bf
p}_2)\psi({\bf p}_1)$
 with form factor $\psi({\bf p})=\left(d^2+p^2\right)^{-\frac{1}{4}}$.
The parameter $\alpha$, which determines the asymptotic behavior
of the form factor, is $1/2$, and the relevant solution to the
Lippmann-Schwinger equation,$ t(z)=\left (\lambda^{-1}
-J(z)\right)^{-1}, $ where $ J(z)=\int d^3k\frac{|\psi({\bf
{k}})|^2}{z-E_k}$ has an ultraviolet logarithmic singularity. A
dimensional regularization was used in [6] to render the
Lippmann-Schwinger equation meaningful. In momentum space of
dimension ${\cal D}=3-\varepsilon$, the solution to this equation
has the form
\begin{equation}
\left [t_{\varepsilon}(z)\right ]^{-1}=
\lambda_{\varepsilon}^{-1}-J_{\varepsilon}(z),
\end{equation}
where $J_{\varepsilon}(z)=\int d^{3-\varepsilon}k\frac {|\psi({\bf
{k}})|^2}{z-E_k}$. Prior to making $\varepsilon$ tend to zero, it
is necessary to renormalize the coupling constant. For the
${}^3S_1$ channel of the $NN$ system, the value of the coupling
constant $\lambda_{\varepsilon}$ must be chosen in such a way as
to ensure the existence of a bound-state at $E_B=-2.2246$ MeV in
this channel, in which case the $T$ matrix must have a pole at
$z=E_B$.
\begin{figure}
\resizebox{1\columnwidth}{!}{\includegraphics{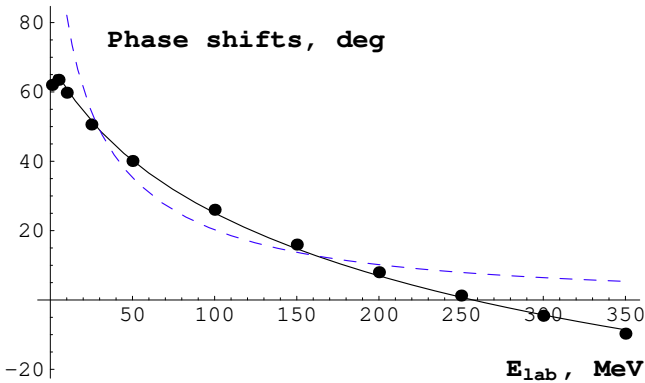}}
\caption{Phase shifts for proton-neutron scattering as a function
of the laboratory energy for the ${}^1S_0$ channel. Experimental
data from [8] are shown by points. The results of the calculation
with the generalized interaction operator (24) are represented by
the solid curve. Also given for the sake of comparison are the
results of the calculation with the Yamaguchi potential (dashed
curve).}
\end{figure}
\begin{figure}
\resizebox{1\columnwidth}{!}{\includegraphics{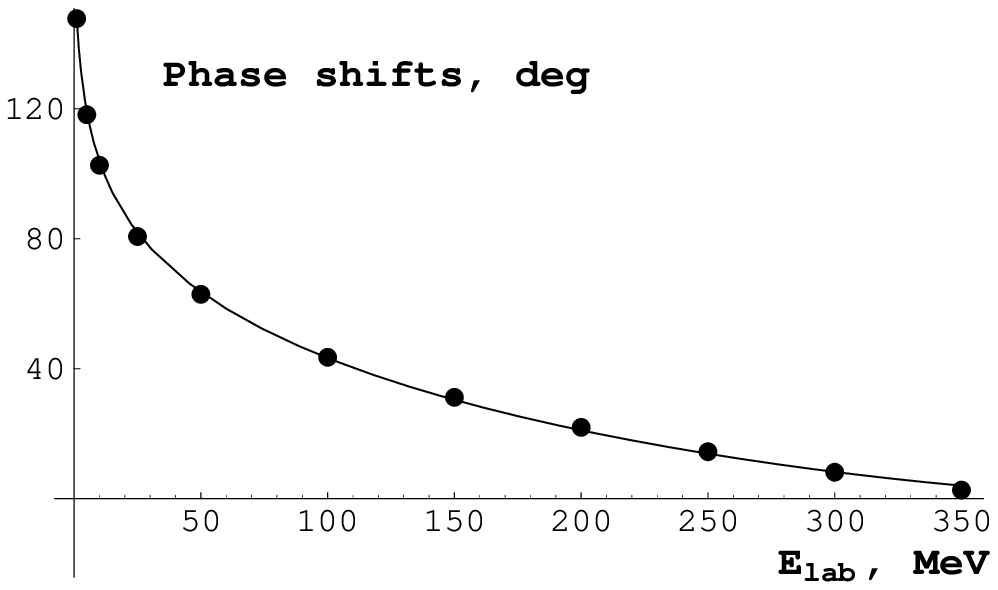}}
\caption{Phase shifts for proton-neutron scattering as a function
of the laboratory energy for the ${}^3S_1$ channel. Experimental
data from [8] are shown by points. The results of the calculation
with the generalized interaction operator (24) are represented by
the solid curve.}
\end{figure}
Thus, the following condition must be satisfied:
\begin{equation}
\lambda_{\varepsilon}^{-1}=J_{\varepsilon}(E_B).
\end{equation}
 The substitution of (29) into (28) then yields
\begin{eqnarray}
[t_{\varepsilon}(z)]^{-1}=J_{\varepsilon}(E_B)-J_{\varepsilon}(z)
=\nonumber\\
= (z-E_B)\int d^{3-\varepsilon}k\frac {|\psi({\bf
{k}})|^2}{(z-E_k)(E_B-E_k)}.\nonumber
\end{eqnarray}
It can easily be proven that, upon going over to the limit
$\varepsilon\rightarrow 0$ in this expression, we arrive at
formula (23), which was previously obtained for the $T$ matrix;
that is, the above renormalization procedure leads to the dynamics
described by the interaction that is specified by Eq. (24) and
which is nonlocal in time. Let us now consider this situation from
the point of view of Hamiltonian dynamics. In the limit
$\varepsilon\rightarrow 0$, the renormalized coupling constant
$\lambda_\varepsilon$  and, hence, the renormalized Hamiltonian
tend to zero, while the $T$ matrix (25) does not satisfy the
Lippmann-Schwinger equation; therefore, the dynamics in question
is not described by the Schr{\"o}dinger equation. Thus, we see
that, although each of the set of dynamics that correspond to the
dimensionality  ${\cal D}=3-\varepsilon$ for $\varepsilon>0$ is a
Hamiltonian dynamics, we have a non-Hamiltonian dynamics in the
limiting case ${\cal D}=3$.  This situation is typical of any
theory where a renormalization procedure is required to remove
ultraviolet divergences. At the same time, the $T$ matrix
satisfies Eq. (7), which one of the possible forms of the master
dynamical Eq. (4) of generalized quantum dynamics. But Eq. (7)
reduces to Lippmann-Schwinger equation only in the particular case
where $H_{int}(t_2,t_1)$ has the form (11)- that is, in the case
of an instantaneous interaction. The dynamics of a renormalized
theory is nonlocal in time; that is, it belongs to the class of
dynamics that can be described only on the basis of generalized
quantum dynamics. For the renormalized model considered here, the
generalized interaction operator is given by (24). This operator
can then be used in a dynamical equation to describe the dynamics
of systems featuring an arbitrary number of nucleons.
\begin{table}
\caption{The parameters of the interaction operator obtained  by
fitting the $NN$ date,  $\rho=1MeV^{-1}$.}
\begin{tabular}{|c|c|c|c|c|c|}
\tableline partial wave&  $\alpha$ &$\lambda$ & $b\cdot \rho$ &
$d\cdot \rho$ & $ b_2\cdot \rho^{1-2\alpha}$ \\
\tableline ${}^3S_1($np$)$ & 0.499 & $133.5\times 10^2$ &433.8 &
$766.2$
& $1.696\times 10^{-7}$  \\
\tableline ${}^1S_0($np$)$ & 0.499 & 131.8 & 356.3 & $3.651\times
10^6$  & $1.694\times 10^{-7}$ \\
\tableline ${}^1S_0($pp$)$ & 0.499 & 320.0 & 371.7 & $6.763\times
10^5$  & $1.695\times 10^{-7}$ \\
\tableline
\end{tabular}
\end{table}

\section{Anomalous off-shell behavior of two-nucleon amplitudes}

Thus, we have shown that the regularization of the Schr{\"o}dinger
and Lippmann-Schwinger equations, which is necessary in using
effective interaction operators constructed within effective field
theories results in that the interaction generating nucleon
dynamics appears to be nonlocal in time. The evolution of systems
governed by such interactions is described in a natural way, by
generalized quantum dynamics and by models constructed on its
basis [4,5]. In [5], the $NN$ interaction was described on the
basis  of the model where the generalized interaction operator has
the form (24) with form factor $ \psi({\bf p})= g_Y({\bf
p})-\phi({\bf p})$, where $g_Y({\bf p})$ is the Yamaguchi form
factor [7], which, in the S channel, is given by $g_Y({\bf
p})=\frac{\lambda}{b^2+p^2}$, and  $\phi({\bf p})=\left(d^2+
p^2\right)^{-\frac{\alpha}{2}}$ with
$-\frac{1}{2}<\alpha<\frac{1}{2}$; that is, $\phi({\bf p})$ is the
form factor whose ultraviolet behavior corresponds to an
interaction that is nonlocal in time. The parameters of the model
were determined from the best fit to the experimental values [8]
of the phase shift for nucleon-nucleon scattering at low energies.
For the ${}^1S_0$ and the ${}^3S_1$ channel, the quality of our
fits to the experimental values of the phase shifts for
nucleon-nucleon scattering are illustrated in Figs. 1-3. The
parameters of the model are quoted in the table. For the sake of
comparison, the energy dependence of the phase shift for
nucleon-nucleon scattering is also displayed in Fig. 1. From this
figure, it can be seen that, in the Yamaguchi model, the main
flaw, which consists in its inability to reproduce the reversal of
the sign of the phase shift in the ${}^1S_0$ channel can be
removed by generalizing this model to the case where the
interaction is nonlocal in time.
\begin{figure}
\resizebox{1\columnwidth}{!}{\includegraphics{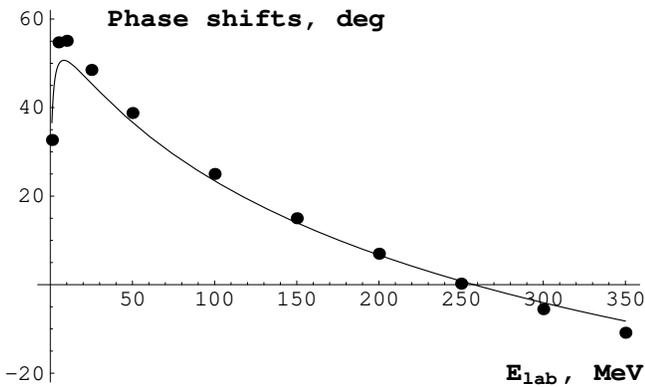}}
\caption{Phase shifts for proton-proton scattering as a function
of the laboratory energy for the ${}^1S_0$ channel. Experimental
data from [8] are shown by points. The results of the calculation
with the generalized interaction operator (24) are represented by
the solid curve.}
\end{figure}

Needless, to say, the $NN$-interaction potential constructed in
this study is nothing but a model-dependent quantity. A realistic
effective $NN$-interaction operator that takes into account QCD
symmetries must be derived within an effective field theory by
using a renormalization procedures. However, our exactly solvable
model can be employed the study the effect of the nonlocality of
$NN$ interaction on the character of nucleon dynamics. Among data
from two-nucleon physics, information about the off-shell of
two-nucleon amplitudes is of great importance, since it
substantially affects the dynamics of three-nucleon and
multinucleon systems [9]. Let us consider the effect of
nonlocality of the $NN$ interaction on this behavior of
two-nucleon amplitudes. First of all, we consider the behavior of
$f(z)=\langle{\bf p}_2|T(z)|{\bf p}_1\rangle$ as a function $z$ at
fixed ${\bf p}_1$ and ${\bf p}_2$. It is well known that, for
$|z|\to \infty$, the solutions $\langle{\bf p}_2|T(z)|{\bf
p}_1\rangle$ to the Lippmann-Schwinger equation tend to
$\langle{\bf p}_2|V|{\bf p}_1\rangle$, where $V$ is the potential.
Thus, we see that, in the case where the $NN$ interaction is
described by some potential-that is, this interaction is local in
time - the two-nucleon amplitude $\langle{\bf p}_2|T(z)|{\bf
p}_1\rangle$ fends to a nonzero constant for $|z|\to \infty$. At
the same time, $\langle{\bf p}_2|T(z)|{\bf p}_1\rangle$ taken at
fixed ${\bf p}_1$ and ${\bf p}_2$ always tends to zero for $|z|\to
\infty$ in the case of an interaction that is nonlocal in time.
Indeed, we have already indicated that, in the nonlocal case,
$H_{int}^{(s)}(\tau)$ does not have a delta-function singularity
at the point $\tau=0$, whence one can immediately conclude that
$B(z)$, which is defined by the relation (10), tends to zero for
$|z|\to \infty$. According to (9), it immediately follows that, in
this limit, $\langle{\bf p}_2|T(z)|{\bf p}_1\rangle$ also tends to
zero. For our nonlocal model, as well as for Yamaguchi model, Fig.
4 illustrates the behavior of the function $f(z)$ in the
${}^3S_1(np)$ channel. It is obvious that this anomalous behavior
of two-nucleon amplitudes, which is due to the nonlocality of the
$NN$ interaction in time, can significantly affect the dynamics of
multinucleon systems.

As was shown above, the generalized interaction operator can be
nonlocal in time only if its matrix elements $\langle{\bf
p}_2|H_{int}^{(s)}(\tau)|{\bf p}_1\rangle$ as functions of momenta
have an ultraviolet behavior that leads to divergences in
Hamiltonian dynamics. Accordingly, the $T$-matrix elements
$\langle{\bf p}_2|T(z)|{\bf p}_1\rangle$ as functions of ${\bf
p}_1$ and ${\bf p}_2$ will not decrease at infinity as fast as is
required in Hamiltonian dynamics. This brings about the question
of how this circumstance can affect the character of nucleon
dynamics. The importance of the off-shell behavior of the
two-nucleon $T$ matrix is associated with the fact that it appears
in the Faddeev equation, which makes it possible to determine the
$T$ matrix for the system of three nucleons if the two-nucleon $T$
matrix is known. It can straightforwardly be shown, however, that,
if $\langle{\bf p}_2|T(z)|{\bf p}_1\rangle$ does not decrease
sufficiently fast in the high-momentum limit, then the Schmidt
norm for the kernel of the Faddeev equation does not exist at any
value of $z$. Thus, we see that, in the case of an interaction
that is nonlocal in time, the off-shell behavior of two-nucleon
amplitudes is anomalous, which results in that the Faddeev
equation is not well-defined. That effective field theories lead
to a Faddeev equation whose kernel decrease at infinity
insufficiently fast for this equation to be well-defined is one of
the most serious problems in such theories [6]. It is important
that within generalized quantum dynamics the Faddeev equation, as
well as the Lippmann-Schwinger equation, need not be valid, as
this does indeed occur in the case of a nonlocal interaction. One
must then directly use the dynamical Eq. (4) or (7).
\begin{figure}
\resizebox{1\columnwidth}{!} {\includegraphics{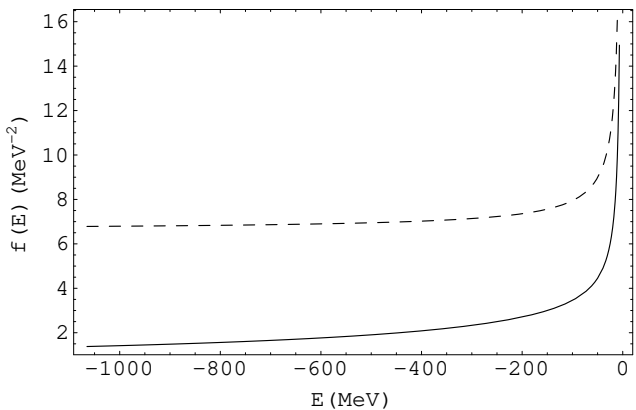}} \caption{The
off-shell behavior of $f(E)=10^9 \langle{\bf p}|T(z)|{\bf
p}\rangle,$ $|{\bf p}|=500$ MeV in the ${}^3S_1$ channel for $np$
scattering. The solid curves corresponds to the model with
generalized interaction operator (24), compared to the model with
Yamaguchi potential with parameters given in \cite{Yam} (dashed
line).}
\end{figure}

In our above analysis, we have considered the case where the
interaction is nonlocal in time and is described by the
interaction operator in the form (24). At the same time, it was
shown in [5] that the generalized interaction operator may have
the form
\begin{eqnarray}
H_{int}(t_2,t_1)=H_{non}(t_2,t_1)\nonumber\\
-2i\delta(t_2-t_1)H_I(t_1),
\end{eqnarray}
where the first term on the right-hand side, $H_{non}(t_2,t_1)$,
describes the nonlocal part of the interaction, while the second
part describes its instantaneous part. This form of the
interaction operator seems natural in the case of $NN$
interactions. Indeed, it is well known that, at long and
intermediate distances, the $NN$ interaction is well approximated
by realistic $NN$ potentials based on the concept of meson
exchange. This part of the $NN$ interaction is described by the
second term on the right-hand side of (30). At the same time,
there is every reason to believe that a nonlocal interaction
operator offers a natural way to treat the short-range part of the
interaction, where quark and gluon degrees of freedom are expected
to manifest themselves. From the above analysis, it follows that
the asymptotic high-momentum behavior of the matrix elements
$\langle{\bf p}_2|H_{int}(t_2,t_1)|{\bf p}_1\rangle$ of the
interaction operator (30) is controlled by the nonlocal term
$H_{non}(t_2,t_1)$. Even if this term makes a negligible
contribution to two-nucleon phase shifts at low energies, it
changes qualitatively the off-shell behavior of two-nucleon
amplitudes and, hence, affects substantially three-nucleon data.
The highlights the importance of taking into account nonlocality
effects in describing the short-range part of the $NN$
interaction. The use of nonlocal interaction operators for the
short-range part of the $NN$ interaction, along with realistic
$NN$ employed at present, may lead to a better description of
three-nucleon and multinucleon data. We hope that it will be
possible to construct such operators - that is, those nonlocal
interaction operators that would describe the short-range part of
the $NN$ interaction - on the basis of effective field theories.

\section{Conclusion}

By considering the example solvable model, we have shown that,
upon the application of regularization and renormalization
procedures, the dynamics of a nucleon system governed by an
interaction that involves ultraviolet divergences is not
Hamiltonian - it is described by the dynamical Eq. (4) featuring a
generalized interaction operator that is nonlocal in time. Here,
we are dealing with dynamics that can be consistently described
only within generalized quantum dynamics. Thus, generalized
quantum dynamics opens new possibilities for solving problems
associated with the fact that effective field theories lead to
effective nucleon-nucleon interaction operators involving
ultraviolet divergences. It can be expected that nucleon dynamics
to which effective field theories must lead will be described by
some generalized interaction operator that is nonlocal in time.
If, within an effective field theory, one will be able to
construct such an operator, which will then respect QCD
symmetries, it will be possible to use Eq. (4) to describe nucleon
dynamics. For the example of the aforementioned model, we have
shown that such an operator can be constructed. We have
investigated the effect of the nonlocality of $NN$ interaction in
time on the character of nucleon dynamics. Our analysis has
revealed that these effects lead to an anomalous off-shell
behavior of two-nucleon amplitudes: The two-nucleon amplitudes
$\langle{\bf p}_2|T(z)|{\bf p}_1\rangle$ at fixed momenta vanish
for $|z|\to\infty$ and, treated as functions of ${\bf p}_1$ and
${\bf p}_2$, decrease insufficiently fast at infinity for the
Faddeev equation to be well-defined. This may substantially affect
the dynamics of multinucleon systems. As we have shown, the
nonlocal interaction operator constructed here can be used for the
nonlocal part of the $NN$-interaction operator. At the same time,
realistic $NN$ potentials can be taken for its instantaneous part
describing the $NN$ interaction at intermediate and long
distances. The introduction of such nonlocal corrections to
realistic $NN$ potentials may significantly improve the
description of three-nucleon and multinucleon data, which is one
of the challenging problems in nucleon physics.

\section*{Acknowledgments}
We would like to thank W. Scheid for helpful discussions and
valuable comments. R.Kh.G. would like to acknowledge the
hospitality of Institut f{\"u}r Theoretische Physik der
Justus-Leibig-Universit{\"a}t,Giessen, where part of this work was
completed.

This work was supported by the Academy of Sciences of Tatarstan
[grant no. 14-98/2000(F)].


\begin{references}
\bibitem{EFT1}
S. Weinberg,  Phys. Let. B {\bf 251}, 288 (1990); Nucl. Phys. B
{\bf 363}, 3 (1991); Ordonez C. and van Kolck U.,  Phys. Let. B
{\bf 291}, 459 (1992); Ordonez C., Ray L. and van Kolck U.,  Phys.
Rev. Let. {\bf 72}, 1982 (1994); Phys. Rev. C {\bf 53}, 2086
(1996); van Kolck U.,  Phys. Rev. C {\bf 49} 2932 (1994).
\bibitem{2} Yu.S. Kalashnikova, I.M.Narodetsky, and V.P.Yurov,
Yad. Fiz., {\bf 49}, 232 (1989);
 Yu.A.Simonov,  Phys. Lett. B {\bf 107}, 1 (1981);
 A.G. Baryshnikov, L.D.Blokhintsev, I.M.Narodetsky, and D.A.Savin,
 Yad. Fiz. {\bf 48}, 1273 (1988);
  A.N. Safronov, Teor. Mat. Fiz. {\bf 89}, 420 (1991);
 Yad. Fiz., {\bf 57}, 208 (1994);
  Yu.A. Kuperin, K.A. Makarov, and S.P.Merkuriev,
Teor.Mat.Fiz., {\bf 75}, 431 (1988); {\bf 76}, 242 (1989);
 A. Abdurakhmanov and A.L. Zubarev, Z.Phys. {\bf A322}, 523
 (1985); M. Orlowski,  Helv.Phys.Acta. {\bf 56}, 1053 (1983);
B.O. Kerbikov, Yad. Fiz. {\bf 41}, 725 (1985); Teor. Mat. Fiz.
{\bf 65}, 379 (1985).
\bibitem{R.Kh.:1999} R.Kh. Gainutdinov,
J. Phys. A\ {\bf 32}, 5657 (1999).
\bibitem{R.Kh./A.A.:1997} R.Kh. Gainutdinov and  A.A. Mutygullina,
Yad.\ Fiz.\ {\bf 60}, 938 (1997).
\bibitem{R.Kh./A.A.:1999} R.Kh. Gainutdinov and  A.A. Mutygullina,
Yad.\ Fiz.\ {\bf 62}, 2061 (1999).
\bibitem{Phillips:2000} D.R. Phillips, I.R. Afnan,
and A.G. Henry-Edwards,
 Phys.\ Rev. C.\ {\bf 61}, 044002-1 (2000).
\bibitem{Yam}
Y.Yamaguchi,  Phys.\ Rev.\ {\bf 95}, 1635 (1954).
\bibitem{Stoks:1993}
V.G.J. Stoks, R.A.M. Klomp, M.C.M. Rentmeester, and J.J. de Swart,
Phys.\ Rev. C\, {\bf 48}, 792 (1993).
 \bibitem{Machleidt}
 R. Machleidt, F. Sammarruca, and Y. Song, Phys.\ Rev. C\ {\bf 53}, 1483
(1996); A. D. Lahiff and I. R. Afnan, Phys.\ Rev. C\ {\bf 56},
2387 (1997).

\end{references}
\end{document}